\documentclass[11pt,a4paper]{article}

\usepackage{mathpazo}
\usepackage{mathbbol,amsthm,bm,amsmath}
\usepackage{graphicx}
\usepackage{color}
\usepackage{soul}
\usepackage{url}
\usepackage{a4wide}
\usepackage{soul}
\usepackage{rotating}
\usepackage{cite}

\newcommand{\e}[2]{$#1 \times 10^{-#2}$}

\graphicspath{{./}{./docs_anagha/}}

\title{Multi-species network inference improves gene regulatory
  network reconstruction for early embryonic development in
  \textit{Drosophila}}

\author{Anagha Joshi$^1$, Yvonne Beck$^{2,\sharp}$ and Tom Michoel$^{2,\ast}$}

\date{}

\begin{document}

\maketitle

$^1$Division of Developmental Biology and $^2$Division of Genetics and
Genomics, The Roslin Institute, The University of Edinburgh,
Midlothian EH25 9RG, Scotland, United Kingdom

\medskip

$^\sharp$Current address: Institute for Applied System Dynamics,
Aalen University, Beethovenstrasse 1, 73430 Aalen, Germany

\medskip

$^\ast$Corresponding author, E-mail: tom.michoel@roslin.ed.ac.uk

\medskip

\begin{abstract}
  Gene regulatory network inference uses genome-wide transcriptome
  measurements in response to genetic, environmental or dynamic
  perturbations to predict causal regulatory influences between
  genes. We hypothesized that evolution also acts as a suitable
  network perturbation and that integration of data from multiple
  closely related species can lead to improved reconstruction of gene
  regulatory networks. To test this hypothesis, we predicted networks
  from temporal gene expression data for 3,610 genes measured during
  early embryonic development in six \textit{Drosophila} species and
  compared predicted networks to gold standard networks of ChIP-chip
  and ChIP-seq interactions for developmental transcription factors in
  five species. We found that (i) the performance of single-species
  networks was independent of the species where the gold standard was
  measured; (ii) differences between predicted networks reflected the known
  phylogeny and differences in biology between the species; (iii) an
  integrative consensus network which minimized the total number of
  edge gains and losses with respect to all single-species networks
  performed better than any individual network. Our results show that
  in an evolutionarily conserved system, integration of data from
  comparable experiments in multiple species improves the inference of
  gene regulatory networks. They provide a basis for future studies on
  the numerous multi-species gene expression datasets for other
  biological processes available in the literature.
\end{abstract}

\newpage

\section{Introduction}
\label{sec:introduction}

In systems biology it is hypothesized that causal regulatory
influences between transcription factors (TFs) and their target genes
can be reconstructed by observing changes in gene expression levels
during dynamic processes or in response to perturbing the cell by gene
mutations or extra-cellular signals \cite{idek01b,kitano2002}. As
increasing amounts of gene expression data have become available,
numerous computational and statistical methods have been developed to
address the gene network inference problem (reviewed in
\cite{friedman2004, gardner2005, bansal2007, albert2007network,
  emmert2012statistical, marbach2012wisdom}). Spurred by the
observation that different methods applied to the same dataset can
uncover complementary aspects of the underlying regulatory network
\cite{michoel2009b,marbach2010revealing}, it is now firmly established
that community-based methods which integrate predictions from multiple
methods perform better than individual methods
\cite{marbach2012wisdom}. A dimension that has remained unexplored in
gene regulatory network inference is evolution: Does the integration
of data from multiple related species lead to improved network
inference performance? Numerous comparative analyses of gene
expression data from multiple species have been performed
\cite{bergmann2003,stuart2003gene, ihmels2005comparative,
  llinas2006comparative, tirosh2006genetic, wang2009meta, lu2009cross,
  kalinka2010gene, miller2010divergence, rhind2011comparative,
  brawand2011evolution, mutwil2011planet, romero2012comparative,
  movahedi2012comparative, roy2013arboretum,
  thompson2013evolutionary}, but invariably these have studied
conservation and divergence of individual gene expression profiles or
co-expression modules. However, it is known that (co-)expression can
be conserved despite divergence of upstream \textit{cis}-regulatory
sequences, and although shuffling of TF-binding sites does not
necessarily alter the topology of the TF--target network, cases have
been documented where conserved co-expression modules are regulated by
different TFs in different species (``TF switching'') (reviewed in
\cite{weirauch2010conserved}). It is therefore not a priori obvious if
and how multi-species expression data can be harnessed for gene
regulatory network inference.

To address this question we decided to focus on a regulatory model
system that is well characterized and conserved across multiple
species. We were therefore particularly interested in a study where
gene expression was measured at several time points during early
embryonic development in six \textit{Drosophila} species, including
the model organism \textit{D. melanogaster}
\cite{kalinka2010gene}. Early development of the animal body plan is a
highly conserved process, controlled by gene regulatory network
components resistant to evolutionary change
\cite{davidson2006gene}. Furthermore, the binding sites of around half
of all sequence-specific regulators controlling transcription in the
blastoderm in \textit{D. melanogaster} have been mapped on a
genome-wide scale by ChIP-chip \cite{macarthur2009developmental} and
for several of these factors additional binding profiles mapped by
ChIP-sequencing are available in other \textit{Drosophila} species
\cite{bradley2010binding, he2011high, paris2013extensive}. In this
study we took advantage of these unique gold standard networks of
regulatory interactions across multiple species to predict and
evaluate gene regulatory networks from gene expression data in six
species, study their phylogeny and biology, and analyze how an
integrated multi-species approach improves network inference
performance.

\section{Results}
\label{sec:results}

\subsection{Evolutionary and developmental dynamics have comparable
  effects on gene expression}
\label{sec:evol}

We collected gene expression data for 3,610 genes in six
\textit{Drosophila} species measured at 9--13 time points during early
embryonic development with 3--8 replicates per time point (200 samples
in total) \cite{kalinka2010gene}. To obtain a global view on the
similarities and differences between samples, we performed
multi-dimensional scaling using Sammon's nonlinear mapping criterion
on the 3,610-dimensional sample vectors (cf. Methods and Figure
\ref{fig:sammon}a). The first (horizontal) axis of variation
corresponded to developmental time, with samples ordered along this
dimension according to increasing developmental time points, while the
second (vertical) axis of variation corresponded to evolutionary
distance, with samples ordered along this dimension according to
species. By expanding these two axes of variation into principal
components, we found that the ``developmental'' dimension explained
34\% of the total variation in the data, while the ``evolutionary''
dimension explained 11\% (cf. Methods). This result confirms that
variations in gene expression levels across \textit{Drosophila}
species at the same developmental time point are not greater than
variations across time points within the same species. In this study
we were interested whether this additional layer of inter-species
expression variation can be harnessed in the reconstruction of gene
regulatory networks.

\subsection{Single-species network reconstruction recovers known
  transcriptional regulatory interactions in early \textit{Drosophila}
  development}

We used the context-likelihood of relatedness (CLR) algorithm
\cite{faith2007} with Pearson correlation as a similarity measure to
predict regulatory interactions in each species separately from the
developmental gene expression data. As candidate regulators we used a
set of 14 sequence-specific transcription factors (TFs) present on the
expression array whose binding sites have been mapped by ChIP-chip in
\textit{D. melanogaster} at developmental time points relevant for the
present study \cite{macarthur2009developmental}. A gold standard
network of known transcriptional regulatory interactions in
\textit{D. melanogaster} development was constructed by assigning
binding sites of these TFs to their closest gene (cf. Methods).  The
gold standard network was dense (25\% of all possible edges were
present) consistent with the fact that genes on the expression array
were selected from genes known to be expressed during embryonic
development \cite{kalinka2010gene} and that the 14 TFs comprise
one-third of all sequence-specific regulators controlling
transcription in the \textit{D. melanogaster} blastoderm embryo
\cite{macarthur2009developmental}.

\begin{figure}[h!]
  \centering
  \includegraphics[width=\linewidth]{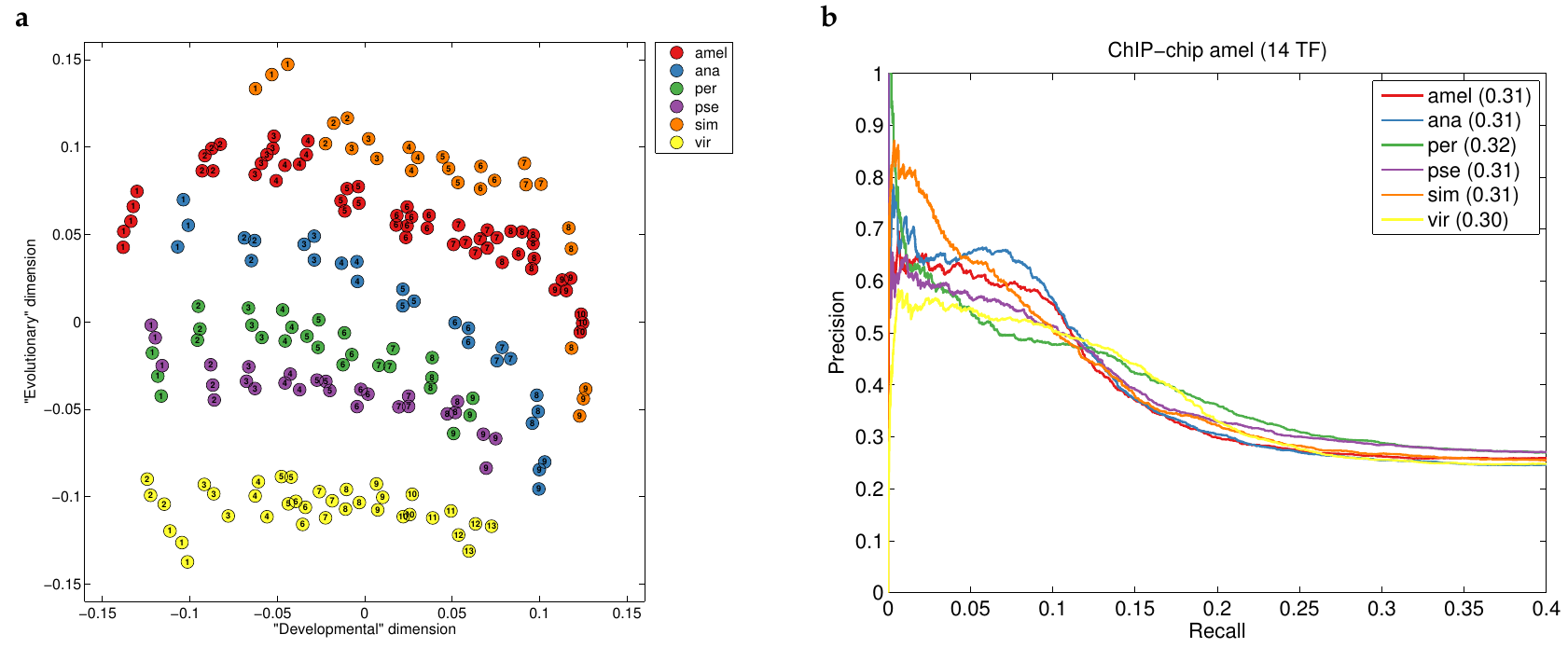}
  \caption{\textbf{a.} Two-dimensional scaling plot of the gene
    expression data using Sammon's nonlinear mapping criterion. Each
    dot represents one sample (200 samples total) positioned such that
    the two-dimensional distances reflect the Euclidean distances
    between the 3610-dimensional data vectors. Samples are colored by
    species and the number in each dot is the developmental time point
    of the sample. \textbf{b.} Recall vs. precision curves for
    predicted regulatory networks in six \textit{Drosophila} species
    using a gold standard network of ChIP-chip interactions for 14 TFs
    in \textit{D. melanogaster}}
  \label{fig:sammon}
\end{figure}

We compared the predicted regulatory networks in all six species to
the \textit{D. melanogaster} gold standard network using standard
recall and precision measurements
\cite{stolovitzky2009lessons}. Without exception all six predicted
networks showed percentages of true positives close to or in excess of
50\% at a recall level of 10\%, corresponding to networks with
1,300--1,400 predicted interactions (Table \ref{tab:refnet} and Figure
\ref{fig:sammon}b). Any differences in performance between species
were found to be small (nearly identical areas under the curve (AUC),
Figure \ref{fig:sammon}b).  The recall cut-off of 10\% in Table
\ref{tab:refnet} was chosen because it was closest for most species to
the inflection point where precision starts to drop more rapidly with
increasing recall. The levels of accuracy in network prediction
obtained here have previously only been observed for bacteria
\cite{michoel2009b,marbach2012wisdom} and demonstrate the importance
of using a gold standard network measured in an appropriate
experimental condition.  Indeed, when we used the more heterogeneous
modENCODE \cite{roy2010identification} or Flynet
\cite{marbach2012predictive} \textit{D. melanogaster} reference
networks, performance dropped dramatically (data not shown).

\subsection{Chip-sequencing data confirms similar network
  reconstruction performance independent of species}
\label{sec:chip-sequ-refer}

Although the gold standard network reconstructed from ChIP-chip data
was in \textit{D. melanogaster}, perhaps surprisingly the
\textit{D. melanogaster} predicted network did not perform better
overall than the networks predicted for the other species (Figure
\ref{fig:sammon}b). To get confidence in this observation, we
downloaded ChIP-sequencing data for three TFs (BCD, KR, HB) in three
\textit{Drosophila} species (\textit{melanogaster},
\textit{pseudoobscura} and \textit{virilis}) \cite{paris2013extensive}
and one TF (TWI) in four species (\textit{melanogaster},
\textit{simulans}, \textit{ananassae} and \textit{pseudoobscura})
\cite{he2011high}, and created ChIP-seq gold standard networks for
five species (cf. Methods).  The recall-precision curves generated
from the \textit{D. melanogaster} ChIP-seq gold standard network
(Supplementary Figure \ref{fig:rec-prec-single}b) were in good
agreement with the ChIP-chip data, demonstrating again that the
\textit{D. melanogaster} predicted network performed no better than
other \textit{Drosophila} species. We also calculated recall-precision
curves using the \textit{D. ananassae}, \textit{D. pseudobscura},
\textit{D. simulans} and \textit{D. virilis} ChIP-seq gold standard
networks. Again, the regulatory network in that species did not
perform better compared to the other species (Supplementary Figure
\ref{fig:rec-prec-single}c--f).

\subsection{Reconstructed regulatory networks are enriched for
  ubiquitous interactions}
\label{sec:reconstr-regul-netw}

\begin{figure}
  \centering
  \includegraphics[width=0.7\linewidth]{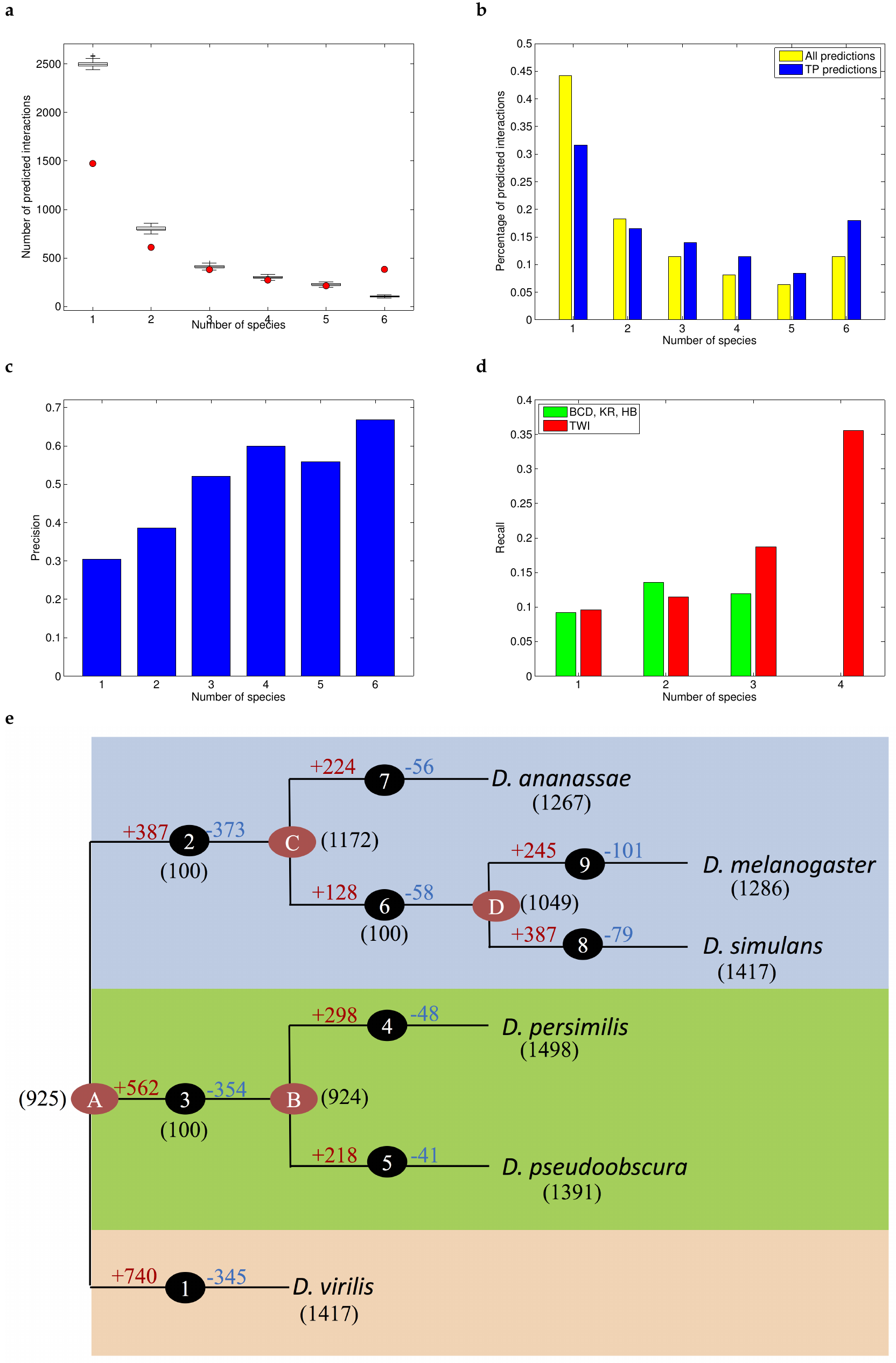}
  \caption{\textbf{a.} Number of interactions found in one to six
    species in the inferred gene regulatory networks at 10\% recall
    level (red dots) and in 100 randomized networks with the same in-
    and out-degree distribution as the inferred networks
    (boxplots). \textbf{b.} Percentage of all predicted interactions
    (yellow) and of all true positive predictions (blue) in one to six
    species \textbf{c.} Precision of interactions found in one to six
    species. \textbf{d.} Recall of ChIP-seq gold standard interactions
    conserved in one to three species (green; data for BCD, KR, HB)
    and one to four species (red; data for TWI). \textbf{e.} Phylogenetic tree
    between six \textit{Drosophila} species reconstructed from the
    inferred interactions at 10\% recall level, with the total number
    of interactions in each species shown in brackets. The tree
    correctly splits the species in 3 groups -- \textit{melanogaster}
    (top), \textit{obscura} (middle), \textit{virilis} (bottom). Each
    branch, (numbered 1--9) represents a inferred network state
    transition. At each network state transition, the number of
    interactions inferred to be gained (red) or lost (blue) as well as
    the bootstrap value for each branch (in brackets) is indicated.}
  \label{fig:phylo}
\end{figure}

The result that network reconstruction performance is similar across
species regardless of the species-origin of the gold standard network
suggests that each species-specific dataset represents a different
perturbation of an underlying conserved regulatory network. To better
understand how the predicted networks in each species relate to each
other, we analysed the reconstructed regulatory networks at the 10\%
recall level (Table \ref{tab:refnet}) in greater detail. Taken
together, these networks contained 3,329 regulatory interactions
between 14 TFs and 1098 genes. About 10\% of these interactions (382)
were predicted in all species. To systematically evaluate if this
overlap can occur by chance, we randomized independently each
interaction network keeping its in- and out-degree distribution
constant and calculated the frequencies of having one to six edges
overlap in 100 randomized networks.  The predicted networks were
significantly enriched for interactions ubiquitous to all species
($Z$-score$=37.7$) and depleted for species-specific interactions
($Z$-score$=-39.5$) (Figure~\ref{fig:phylo}a).

We then calculated if individual TFs were biased towards
species-specific or ubiquitous interactions. Zygotic factors such as
SNA ($P = 9.8\times 10^{-60}$) shared statistically significant
predicted targets among all six species whereas maternal factors such
as CAD did not share a single target across the six species. This
together with the observation that early zygotic genes at sequence
level evolved much slower \cite{mensch2013positive} leads to the
hypothesis that not only the sequences of early zygotic lineage genes
but also the transcriptional program controlling their expression has
evolved slower. The early zygotic genes are indeed overrepresented in
the targets with conserved interactions across all species
($P=1.2\times 10^{-5}$).

The observation that prediction performance is independent of species
(Figure \ref{fig:rec-prec-single}) could be explained if only
ubiquitous interactions (predicted in all species) were true
positives. Although more true positives are found among interactions
shared by three or more species than expected based on the total
distribution of predicted interactions (Figure \ref{fig:phylo}b), and
with precision increasing by the number of species (Figure
\ref{fig:phylo}c), ubiquitous interactions account for only 18\% of
all true positives.  Another possible explanation for the
species-independent performance could be that binding events are
highly conserved across species. Although it has been noted that more
than 90\% of TF binding sites overlapped between
\textit{D. melanogaster} and the closely related \textit{D. yakuba}
\cite{bradley2010binding}, less than 30\% of those binding sites were
also conserved in the more distant \textit{D. pseudoobscura}
\cite{paris2013extensive}. Furthermore it is also not true that
conserved gold standard interactions for these TFs (BCD, HB and KR)
are more likely to be inferred. Indeed, the recall for
species-specific gold standard interactions or those conserved in two
or three species for these factors in the 10\% recall networks did not
differ from the overall recall value (Figure \ref{fig:phylo}d). In
contrast, for the factor TWI, gold standard interactions conserved in
three or four species were more likely to be included in the 10\%
recall networks (recall values resp. 19\% and 36\%, Figure
\ref{fig:phylo}d). This is consistent with a higher degree of binding
site conservation for this factor with up to 60\% conserved binding
sites across six species \cite{he2011high}.

\subsection{Differences between predicted transcriptional regulatory
  networks reflect known phylogeny and biology}
\label{sec:simil-pred-netw}

Since conservation of predicted or known gold standard interactions
across species does not fully explain the observed species-independent
network reconstruction performance, we hypothesized that the
differences between these networks are not solely due to random
variations in the expression data. To analyse these differences, we
constructed a phylogenetic tree between the species based on the gain
or loss of predicted interactions using the principle of maximum
parsimony. This method minimises the number of state changes in all
transitions in a tree and has been used previously to reconstruct the
evolutionary history of species based on gene content
\cite{martens2008whole} and to reconstruct and predict transition
states of developmental lineage trees based on gene expression data
\cite{joshi2011maximum}. Using a binary matrix representing the
presence or absence of all 3,329 predicted TF-target interactions in
each of the 10\% recall networks, a rooted tree was reconstructed
which split the species in three groups: \textit{melanogaster} (top),
\textit{obscura} (middle), \textit{virilis} (bottom) (cf. Methods and
Figure \ref{fig:phylo}d). This tree is in full agreement with the tree
reconstructed based on gene content \cite{stark2007discovery}.  To
ensure the robustness of the tree, we applied a standard bootstrap
procedure which predicted 100\% bootstrap confidence on all branches
of the tree (Figure \ref{fig:phylo}d). The parsimony tree, moreover,
predicts the network state transitions at each branch in terms of
interactions gained or lost at a given transition. The transitions
show a bias towards gain of interactions at most branch points over
the loss. This is probably due to the presence of a large number of
species-specific interactions (Figure \ref{fig:phylo}a).

We further explored whether the nine branch points (numbered 1--9 in
Figure \ref{fig:phylo}d) reflect the biology behind the evolution of
the \textit{Drosophila} species. We created gene lists at each branch
point containing target genes which gained or lost transcriptional
interactions at that branch point. The maximum number of genes (361)
gained interactions from branch point `A' to \textit{D. virilis} and
were enriched for neuron differentiation ($P=1.2\times 10^{-6}$) and
embryonic morphogenesis ($P=3.1\times 10^{-8}$). Genes gaining
interactions from branch point `D' to \textit{D. simulans} were
enriched for response to organic substances ($P=3.4\times 10^{-2}$),
in line with the fact that \textit{D. simulans}, unlike
\textit{D. melanogaster}, lives on diverse rotting, non-sweet
substrates throughout the year \cite{david2007historical}. Gene
ontology analysis of all target sets revealed that many gene sets were
enriched for transcription regulation (Supplementary Table
\ref{tab:trans-enrich}), i.e. transcriptional regulators were more
likely to gain or lose interactions in the network rewiring. At each
branch point, we found TFs losing or gaining interactions more than
expected by chance (Supplementary Table \ref{tab:trans-tf}).  For
instance, SLP1 is predicted to lose its interactions with genes
involved in wing disc formation only in \textit{D. ananassae} while
Dorsal (DL) is predicted to regulate mitochondrial genes only in the
\textit{melanogaster} subgroup. Taken together, a biologically
relevant evolutionary network history can be reconstructed using the
individual predicted regulatory networks in six \textit{Drosophila}
species.

\subsection{Multi-species analysis improves network reconstruction}

It has been observed that different network inference algorithms
applied to the same data uncover complementary aspects of the true
underlying regulatory network \cite{michoel2009b,marbach2010revealing}
and this has formed the basis for integrative approaches which combine
the predictions from multiple algorithms \cite{marbach2012wisdom}. In
our case, since the networks predicted from different species equally
well recover known transcriptional interactions while their
differences reflect known phylogeny and biology, we reasoned that a
multi-species analysis which combines predictions across species
should also lead to a better network reconstruction. To test this
hypothesis we considered several integrative approaches. Firstly, we
combined the expression data from all species into one dataset to
which we again applied the CLR algorithm (``merged data''
method). Secondly, we kept CLR scores from the individual species and
applied rank-aggregation methods to derive an ``intersection'',
``union'' and ``average'' consensus ranking of predicted interactions
(cf. Methods).
Finally, motivated by the phylogenetic tree reconstruction, we also
constructed a consensus ranking as the centroid of the six
species-specific rankings for the cityblock distance, which for
discrete networks corresponds to counting total number of edge gains
and losses between two networks (``centroid'' method, see Methods for
details). 

\begin{figure}
  \centering
  \includegraphics[width=0.8\linewidth]{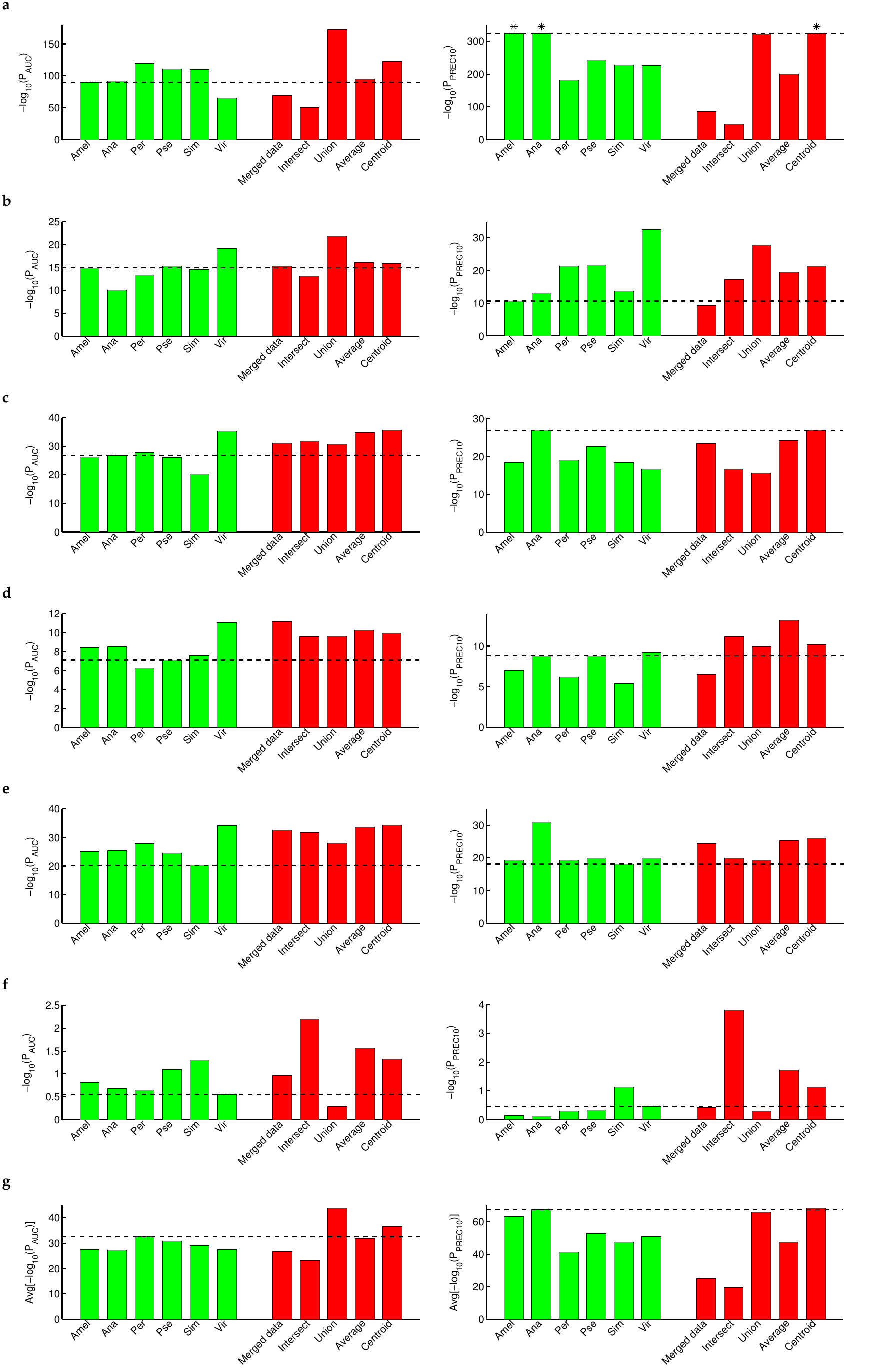}
  \caption{Performance scores with respect to the gold standard
    ChIP-chip network for 14 TFs in \textit{D. melanogaster}
    (\textbf{a}) and the ChIP-seq networks for
    \textit{D. melanogaster} (\textbf{b}, 4 TFs),
    \textit{D. ananassae} (\textbf{c}, 1 TF),
    \textit{D. pseudoobscura} (\textbf{d}, 4 TFs),
    \textit{D. simulans} (\textbf{e}, 1 TF), \textit{D. virilis}
    (\textbf{f}, 4 TFs), and their averages over all gold standard
    networks (\textbf{g}). In each panel, the left, resp. right,
    figure shows $-\log_{10}(P_{\text{AUC}})$,
    resp. $-\log_{10}(P_{\text{PREC10}})$ for the six single-species
    predicted networks (green) and the five prediction aggregation
    methods (red). The dashed lines indicate the performance level of
    the single-species network for the gold standard species
    (\textbf{a--f}) or of the best performing single-species network
    (\textbf{g}). Values with a $\ast$ in panel \textbf{a} indicate
    numerical underflow values truncated to the smallest non-zero
    $P$-value ($10^{-324}$).}
  \label{fig:score}
\end{figure}

To quantitatively compare different methods across different gold
standard networks we considered the area under the recall--precision
curve (AUC) and the precision at 10\% recall (PREC10) as performance
measures and converted them to $P$-values by comparison to AUCs and
PREC10s of networks generated by randomly assigning ranks to all
possible edges in the corresponding gold standard network (cf. Methods
and Figure \ref{fig:rec-prec-aggr} for the recall vs. precision
curves). While the AUC assesses the overall performance of a predicted
network, PREC10 measures the quality of the top-ranked predictions, a
property that may be of greater practical relevance.  This analysis
showed that no predicted network performs best for either measure
across all gold standards (Figure \ref{fig:score}a-f). The
single-species \textit{virilis} networks performed best for 5 out of
12 AUC and PREC10 scores, albeit not for the ChIP-seq network measured
in its own species.  This overall good performance is consistent with
\textit{virilis} having the highest number of measured time points in
the data (Supplementary Table \ref{tab:data}).
\textit{D. melanogaster} also had more data points available than the
other four species, but its time series were less complete
(Supplementary Table \ref{tab:data}). Among the integrative methods,
the centroid and union methods both performed best for 5 out of 12 AUC
and PREC10 scores (Figure \ref{fig:score}a-f). Both also had higher
average AUC score than the best single-species network, but only the
centroid method had higher average PREC10 score than the best
single-species network (Figure \ref{fig:score}g). The most important
result however is the fact that the single-species network for the
species were the gold standard network was measured never has the
highest single-species AUC and only twice has the highest PREC10. In
contrast, the centroid method always performs as good, and in most
cases better, than the single-species network for the reference
species (Figure \ref{fig:score}a-f). We conclude that the centroid
method is the most robust network integration method achieving
consistently high AUC and PREC10 scores, at least on this dataset.

\section{Discussion}
\label{sec:discussion}

Here we predicted and evaluated developmental gene regulatory networks
from temporal gene expression data in six \textit{Drosophila} species,
studied their phylogeny and biology, and analyzed how an integrated
multi-species analysis improved network inference performance using
gold standard networks of regulatory interactions measured by
ChIP-chip and ChIP-seq in five species.

We unexpectedly found that network prediction performance of the
single-species networks was independent of the species where the gold
standard was measured. With precision values around or greater than
50\% at a recall level of 10\% for all predicted networks, this result
was clearly not due to poor overall prediction performance. Although
there was a trend that interactions predicted in all species had
higher precision than interactions predicted in only one species and
that conserved interactions in the gold standard networks for at least
one of the TFs had higher chance to be correctly predicted, neither
trend was sufficiently strong to account for the observed performance
similarities. An alternative or additional explanation could be that
the ``true'' gene expression and binding profiles are highly conserved
between these six species but the observed profiles show
species-dependent variation due to the inherent noisyness of
high-throughput data. Because such random fluctuations in gene
expression and binding data would be unrelated, one would then indeed
expect similar performance independent of species. This explanation
however conflicts with the published findings that binding divergence
for these TFs increases with evolutionary distance and our observation
that the differences between the predicted regulatory networks are
consistent with the known phylogeny and differences in biology between
these six \textit{Drosophila} species. Future work in other species
will have to elucidate if the observed species-independent performance
is an artefact of this particular dataset, a consequence of the highly
conserved nature of the underlying biological process or a more
general feature of this type of analysis.

Motivated by the result that all species-specific networks showed good
inference performance and that their differences reflected true
phylogenetic relations, we pursued integrative approaches whereby
predicted networks from all species were combined into consensus
networks. In addition to established aggregation methods such as
taking the intersection, union or rank average of individual
predictions, we also considered a novel centroid method which
minimizes the total sum of edge gains and losses with respect to all
individual networks. Multi-species methods showed better overall
performance than the single-species networks, consistent with the
observation that correct predictions are not restricted to
interactions predicted in all species. Of note, the single-species
network matching the gold standard species was almost never the best
performing single-species network. Because in real-world applications
the aim of network inference is usually to reconstruct a TF-target
network for a species of interest in the absence of gold standard
ChIP-seq/chip data, our results suggest that by combining predicted
networks from multiple closely related species, a better network will
be inferred than by using data for the species of interest only, and
that the combined network is likely to perform better, or at least as
good as, the best single-species network. A novel multi-species
network integration method which reconstructs an ``ancestral'' network
minimizing the number of edge gains and losses to each single-species
network appeared to be particularly promising in this regard.

Our work has shown that in an evolutionarily conserved system such as
early embryonic development, integration of data from comparable
experiments in multiple species improves the inference of gene
regulatory networks. Although the data for the present study came from
a well-controlled experiment in a model organism, with matching
time-course data adjusted for differences in developmental time
between species, our approach is based solely on comparing expression
profiles of different genes within the same species, and expression
levels in different species were never directly compared. We therefore
expect that our results should also hold for other biological
processes, when more heterogeneous data are used or when data from
more distantly related species are combined, in order to cover the
entire spectrum of available multi-species gene expression datasets.

\section{Methods}
\label{sec:methods}

\subsection{Gene expression data}
\label{sec:gene-expression-data}

Embryonic developmental time-course expression data in 6
\textit{Drosophila} species (\textit{D. melanogaster} (``amel''),
\textit{D. ananassae} (``ana''), \textit{D. persimilis} (``per''),
\textit{D. pseudoobscura} (``pse''), \textit{D. simulans} (``sim'')
and \textit{D. virilis} (``vir'')) was obtained from
\cite{kalinka2010gene} (ArrayExpress accession code E-MTAB-404). The
data consists of 10 (amel), 13 (vir) or 9 (ana, per, pse, sim)
developmental time points with several replicates per time point
resulting in a total of 56 (amel), 36 (vir) or 27 (ana, per, pse, sim)
arrays per species (Supplementary Table \ref{tab:data}). The
downloaded data was processed by averaging absolute expression levels
over all reporters for a gene followed by taking the $\log_2$
transform.

\subsection{Multi-dimensional scaling and variance explained}
\label{sec:multi-dimens-scal}

We used two-dimensional scaling using the Eucledian distance and
Sammon's nonlinear mapping criterion on the 3,610-dimensional sample
vectors using the built-in ``mdscale'' function of Matlab. To estimate
the variance explained by each of the two dimensions, we first
calculated the principal components of the data matrix. These are a
set of 200 mutually orthogonal $(200\times 1)$-dimensional vectors,
each explaining a proportion $\sigma_i^2$ of the total variance, i.e.
$\sum_{i=1}^{200}\sigma_i^2 = 1$. Each dimension in Figure
\ref{fig:sammon} also corresponds to a $(200\times 1)$ vector $Y$ and
the proportion of variance explained by $Y$ is found by expansion into
principal components, $ \sigma^2_Y = \sum_{i=1}^{200} \sigma_i^2
(Y^TV_i)^2$, where it is assumed that $Y$ and all $V_i$ have unit
norm. To correct for systematic biases in the data, genes were
standardized to have mean zero and standard deviation one over all 200
samples.

\subsection{ChIP-chip data}
\label{sec:chip-chip-data}

ChIP-chip data for 21 sequence-specific \textit{Drosophila}
transcription factors (TFs) measured in \textit{D. melanogaster}
embryos was obtained from \cite{macarthur2009developmental}.  We
considered the 1\% FDR bound regions and defined target genes for each
TF by assigning to each bound region its closest gene, if the distance
between the region and the gene was less than 5,000 base pairs. For
TFs with repeat measurements, target lists were defined by taking the
union over replicates. Fourteen of the TFs were present on the array
and used to construct a gold standard regulatory network.

\subsection{ChIP-sequencing data}
\label{sec:chip-sequencing-data}

The peaks for three transcription factors present on the array (BCD,
HB and KR) for three species (\textit{D. melanogaster},
\textit{D. pseudoobscura} and \textit{D. virilis}) were obtained from
\cite{paris2013extensive}. Genes with normalized peak height greater
than 0 were selected as the gold standard targets of a given
transcription factor. The peaks for one factor (TWI) for four species
(\textit{D. melanogaster}, \textit{D. ananassae},
\textit{D. pseudoobscura} and \textit{D. simulans}) were obtained from
\cite{he2011high}. Peaks were mapped to the nearest transcription
start site of genes by using the gene annotation from FlyBase
(FB2013\_03). Genes with peak height greater than 10 were selected as
the gold standard targets for each species.

\subsection{Transcriptional regulatory network reconstruction}
\label{sec:transcr-regul-netw}

We used the CLR (Context Likelihood of Relatedness) algorithm
\cite{faith2007} using Pearson correlation as a similarity measure to
predict transcriptional regulatory networks in each species, using the
aforementioned 14 TFs as candidate regulators. Because the CLR
algorithm only considers the right-hand tail of similarity values for
every TF--gene combination, in theory the absolute values of the
Pearson correlations should be provided to the CLR software. However,
we observed improved performance with respect to all gold standard
networks when the Pearson correlations were \emph{not} transformed to
absolute values before calling the CLR algorithm (effectively ignoring
negative correlations) and therefore used this approach for all
reported results. Pearson correlation followed by CLR also performed
better than the default mutual information similarity measure followed
by CLR as well as using Pearson correlation or mutual
information without CLR (data not shown).

\subsection{Phylogenetic tree construction}
\label{sec:hier-clust-spec}

We created a binary matrix of 3,329 rows and 6 columns representing
predicted TF--target interactions in each species at a CLR $Z$-score
cutoff corresponding to 10\% recall with respect to the
\textit{D. melanogaster} ChIP-chip network. In this matrix, the
$(i,j)^{\text{th}}$ element denotes whether the interaction $i$ is
present in the species $j$ or not. Network states and state changes
were mapped onto the branches of inferred phylogenetic trees using the
PARS program from the PHYLIP package \cite{felsenstein1996inferring}
by defining \textit{D. virilis} as the root of the tree. Bootstrapping
was performed using the SEQBOOT program from the PHYLIP package where
100 datasets were generated by randomly replacing a given six species
network matrix. A consensus tree with a bootstrap confidence on each
branch of the tree was reconstructed using the CONSENSE program from
the PHYLIP package.

\subsection{Enrichment analyses}
\label{sec:enrichment-analyses}

Gene set enrichment for each phylogenetic tree state change was
calculated using the DAVID suite of programs
\cite{da2008systematic}. For each transcription factor, enrichment of
overlap of the candidate target gene set with each transition state
gene set was calculated using a hypergeometric test.  Early zygotic,
late zygotic, maternal and adult gene lists were downloaded from
\cite{mensch2013positive} and enrichment was calculated using a
hypergeometric test.

\subsection{Prediction aggregation methods}
\label{sec:pred-aggr-meth}

To combine predicted networks from multiple-species, we considered
five prediction aggregation methods. The first method combined
expression data from all species into one dataset to which we again
applied the CLR method (``merged data'' method). For the four other
methods, predictions from each species were first ranked by their
respective CLR-scores, such that the highest score received the
highest rank value and tied values were given their average rank
value, using Matlab's ``tiedrank'' function. Three methods used
standard functions to combine the edge ranks of the six single-species
predicted networks, namely the \emph{minimum} (``intersection''
method), \emph{maximum} (``union'' method) and \emph{average}
(``average'' method) rank value. The intersection and union methods
are named such because if a threshold is used to convert a fixed
number of top-ranked predictions to a binary graph, these would result
precisely in the intersection and union of the binary graphs over all
species. Conversely, on binary graphs, the phylogenetic tree
construction infers ancestral networks by minimizing the number of
edge gains and losses between single-species networks. This
corresponds to minimizing their ``cityblock'' distance, defined for
two $n$-dimensional vectors $x$ and $y$ as $ d(x,y) = \sum_{i=1}^n
|x_i - y_i|$. As the final prediction aggregation method we therefore
considered the centroid network for the six single-species networks
for the cityblock distance, defined by the edge weights $C_{ij}$ which
minimize
\begin{align*}
  \sum_{k=1}^6 d\bigl(C,R^{(k)}\bigr) = \sum_{k=1}^6 \sum_{ij}
  \bigl|C_{ij} - R_{ij}^{(k)}\bigr| 
\end{align*}
where $R^{(k)}$ is the matrix of edge rank values for species $k$. The
matrix $C$ is easily computed using Matlab's ``kmeans'' function, by
specifying the cityblock distance and grouping species into one cluster.

\subsection{Network reconstruction performance}
\label{sec:netw-reconstr-perf}

To compare the network reconstruction performance of several predicted
networks across multiple gold standard networks, we used the area
under the precision-recall curve (AUC) and the precision at 10\%
recall (PREC10) as scoring measures. Absolute scores were converted to
$P$-values following established protocols of the DREAM project
\cite{marbach2012wisdom}. Briefly, 100,000 random predictions were
  generated for each gold standard network by assigning a random rank
  to each possible TF-target interaction. Next, an asymmetric
  stretched exponential function of the form
\begin{align*}
  f(x) =
  \begin{cases}
    h\, e^{-b_+ (x-x_{\max})^{c_+}} & x\geq x_{\max} \\
    h\, e^{-b_- (x_{\max} -x)^{c_-}} & x< x_{\max} 
  \end{cases}
\end{align*}
was fitted to each histogram (1000 bins) of random AUCs and random
PREC10s, using the ``fit'' function in Matlab's Curve Fitting
Toolbox. Finally, $P$-values for real AUCs and PREC10s were calculated
by integrating the right tail of the corresponding fitted and
normalized stretched exponential distribution function using Matlab's
``trapz'' function.

\newpage

\section*{Figures}
\label{sec:figures}

\newpage

\section*{Tables}
\label{sec:tables}

\begin{table*}[h!]
  \centering
  \begin{tabular}{|lrrlrlrlrlrlrl|}
    \hline
    \textbf{TF} & \textbf{ChIP} & \multicolumn{2}{c}{\textbf{Amel}} &
    \multicolumn{2}{c}{\textbf{Ana}} &\multicolumn{2}{c}{\textbf{Per}}
    & \multicolumn{2}{c}{\textbf{Pse}} &
    \multicolumn{2}{c}{\textbf{Sim}} &
    \multicolumn{2}{c|}{\textbf{Vir}} \\ 
    \hline
    D & 1166 & 158 & (129) & 145 & (122) & 171 & (137) & 154 & (124) & 163 & (132) & 132 & (102) \\ 
    kr & 518 & 125 & (86) & 128 & (86) & 196 & (125) & 176 & (109) & 127 & (80) & 207 & (143) \\ 
    mad & 40 & 11 & (0) & 0 & (0) & 1 & (0) & 4 & (0) & 0 & (0) & 0 & (0) \\ 
    bcd & 157 & 13 & (0) & 4 & (0) & 0 & (0) & 0 & (0) & 0 & (0) & 0 & (0) \\ 
    cad & 274 & 8 & (0) & 0 & (0) & 40 & (7) & 0 & (0) & 133 & (7) & 85 & (13) \\ 
    da & 795 & 0 & (0) & 0 & (0) & 0 & (0) & 0 & (0) & 0 & (0) & 0 & (0) \\ 
    dl & 1503 & 216 & (163) & 234 & (183) & 67 & (52) & 137 & (110) & 289 & (216) & 111 & (83) \\ 
    hb & 358 & 0 & (0) & 0 & (0) & 0 & (0) & 0 & (0) & 0 & (0) & 0 & (0) \\ 
    hkb & 206 & 131 & (49) & 181 & (61) & 167 & (45) & 172 & (48) & 135 & (43) & 122 & (34) \\ 
    prd & 313 & 44 & (21) & 38 & (15) & 65 & (28) & 58 & (27) & 41 & (10) & 55 & (22) \\ 
    run & 158 & 134 & (52) & 117 & (49) & 186 & (56) & 154 & (56) & 127 & (47) & 167 & (62) \\ 
    slp1 & 212 & 178 & (57) & 155 & (45) & 221 & (62) & 192 & (57) & 154 & (47) & 192 & (54) \\ 
    sna & 291 & 170 & (78) & 169 & (73) & 207 & (83) & 191 & (76) & 174 & (72) & 197 & (81) \\ 
    twi & 1163 & 98 & (80) & 96 & (81) & 177 & (120) & 153 & (108) & 74 & (61) & 149 & (121) \\
    \hline
    \textbf{Total} & 7154 & \multicolumn{2}{c}{1286} & \multicolumn{2}{c}{1267} &
    \multicolumn{2}{c}{1498} & \multicolumn{2}{c}{1391} &  \multicolumn{2}{c}{1417} &
    \multicolumn{2}{c|}{1417} \\
    \multicolumn{2}{|l}{\textbf{Precision}} & \multicolumn{2}{c}{0.56}
    & \multicolumn{2}{c}{0.56} & \multicolumn{2}{c}{0.48} &
    \multicolumn{2}{c}{0.51} & \multicolumn{2}{c}{0.50} &
    \multicolumn{2}{c|}{0.50}\\ 
    \hline
  \end{tabular}
  \caption{Transcription factors and their number of target genes in
    the \textit{D. melanogaster} ChIP-chip gold standard network and in the
    predicted networks for six \textit{Drosophila} species at the 10\%
    recall level (in brackets for each TF the number of true positive
    predictions). The bottom two rows are the total number of 
    interactions in each network and the overall precision (percentage
    of true positives) of the predicted networks.}
  \label{tab:refnet}
\end{table*}

\newpage

\appendix

\setcounter{equation}{0}
\renewcommand{\theequation}{S\arabic{equation}}
\setcounter{figure}{0}
\renewcommand{\thefigure}{S\arabic{figure}}
\setcounter{table}{0}
\renewcommand{\thetable}{S\arabic{table}}

\section{Supplementary figures}
\label{sec:suppl-figur}

\begin{figure}
  \centering
  \includegraphics[width=\linewidth]{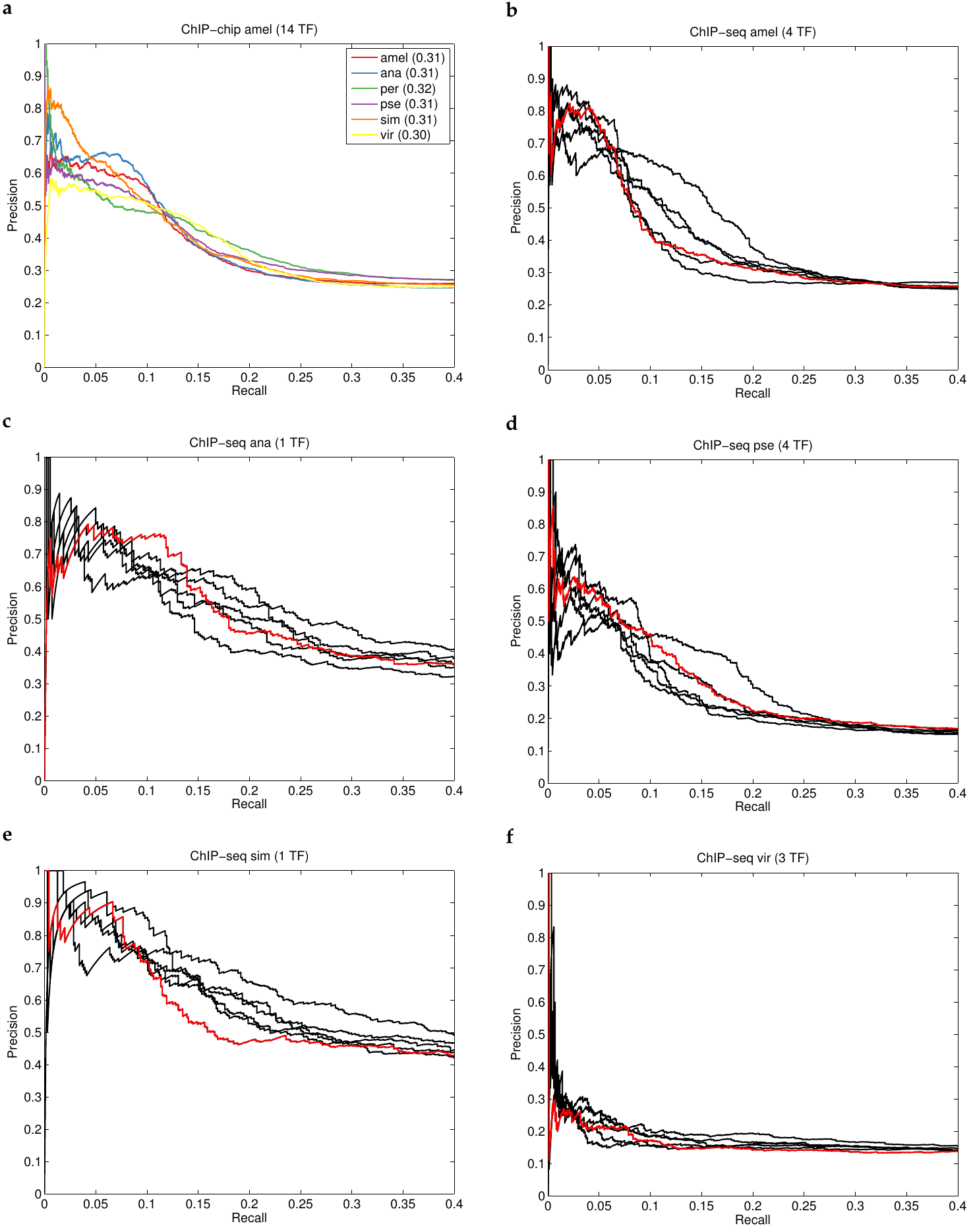}
  \caption{Recall vs. precision curves for predicted regulatory
    networks in six \textit{Drosophila} species. The gold standard
    networks were the ChIP-chip network for 14 TFs in
    \textit{D. melanogaster} (\textbf{a}) and the ChIP-seq networks
    for \textit{D. melanogaster} (\textbf{b}, 4 TFs),
    \textit{D. ananassae} (\textbf{c}, 1 TF),
    \textit{D. pseudoobscura} (\textbf{d}, 4 TFs),
    \textit{D. simulans} (\textbf{e}, 1 TF) and \textit{D. virilis}
    (\textbf{f}, 4 TFs). In panel \textbf{a}, the numbers in the
    legend are the area under the curve for each species. In panel
    \textbf{b--f}, the curve for the reference species is in red while
    the other species are in black.}
  \label{fig:rec-prec-single}
\end{figure}

\begin{figure}
  \centering
  \includegraphics[width=\linewidth]{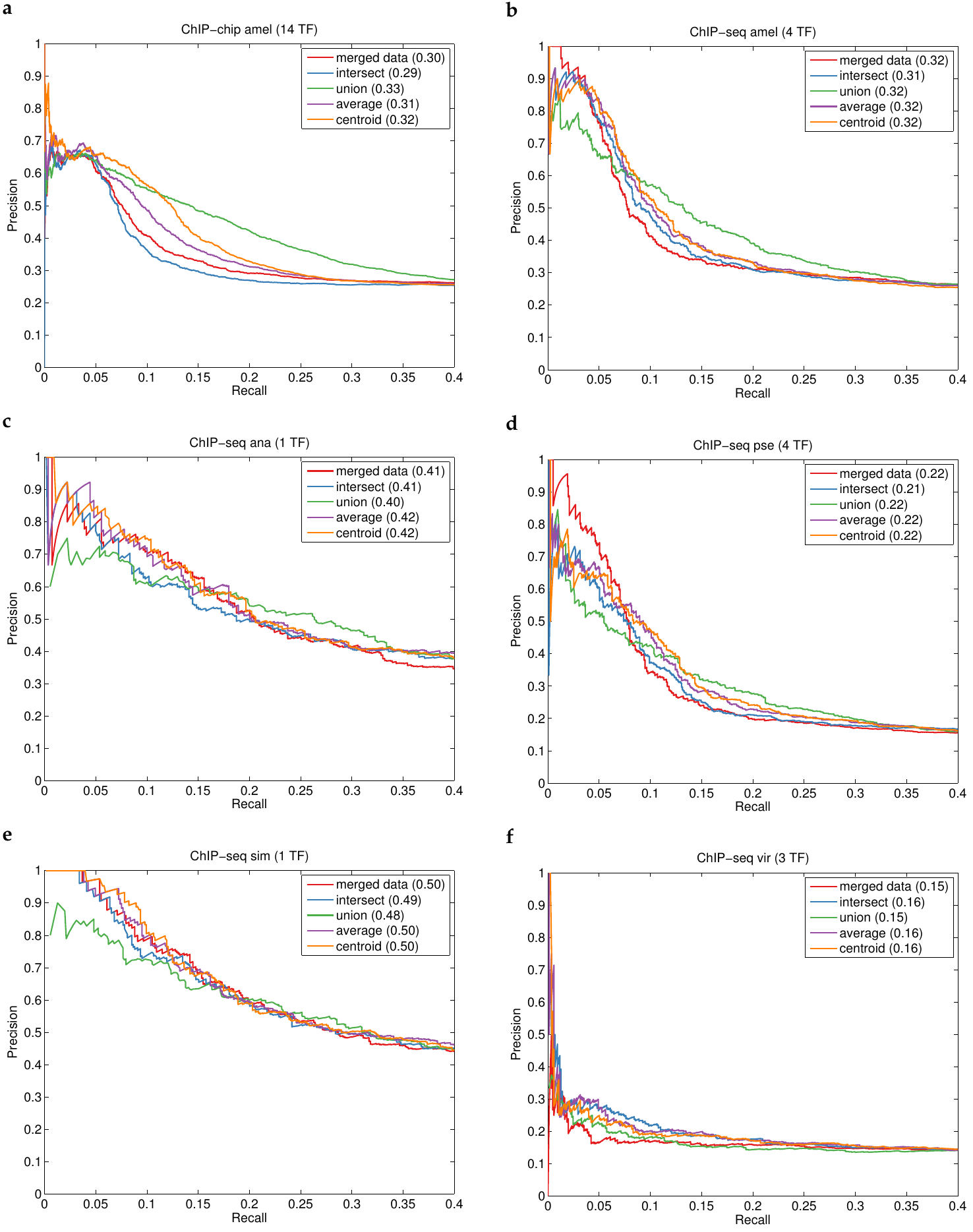}
  \caption{Recall vs. precision curves for predicted regulatory
    networks for five multi-species meta-analysis methods. The gold
    standard networks were the ChIP-chip network for 14 TFs in
    \textit{D. melanogaster} (\textbf{a}) and the ChIP-seq networks
    for \textit{D. melanogaster} (\textbf{b}, 4 TFs),
    \textit{D. ananassae} (\textbf{c}, 1 TF),
    \textit{D. pseudoobscura} (\textbf{d}, 4 TFs),
    \textit{D. simulans} (\textbf{e}, 1 TF) and \textit{D. virilis}
    (\textbf{f}, 4 TFs). The numbers in each legend are the area under
    the curve for each method.}
  \label{fig:rec-prec-aggr}
\end{figure}

\newpage

\section{Supplementary tables}
\label{sec:supplementary-tables}

\begin{table}[h!]
  \centering
  \begin{tabular}{|l|l|l|}
    \hline
    \textbf{Transition} & \textbf{Functional category} &
    \textbf{\textit{P}-value}\\
    \hline
    A $\to$ B loss & post-embyonic organ development &  \e{5.8}{5}\\
    & regulation of transcription & \e{1.2}{4}\\
    A $\to$ B gain & cell fate commitment & \e{1.0}{9}\\
    & regulation of transcription & \e{4.4}{8}\\
    A $\to$ C loss & cell--cell adhesion & \e{2.0}{4}\\
    & exocrine system development &\e{5.9}{4}\\
    A $\to$ C gain & cell fate commitment & \e{1.1}{12}\\
    & regulation of transcription & \e{2.2}{7}\\
    A $\to$ vir loss & regulation of transcription & \e{2.1}{5}\\
    & ectoderm development & \e{4.2}{5}\\
    A $\to$ vir gain & neuron differentiation & \e{1.2}{6}\\
    B $\to$ per loss & positive regulation of apoptosis & \e{6.6}{3}
    \\
    B $\to$ per gain & translation factor activity & \e{2.3}{5}\\
    & regulation of transcription & \e{2.4}{4}\\
    B $\to$ pse loss & sensory organ development & \e{4.6}{3}\\
    & transcription factor activity & \e{8.2}{3}\\
    B $\to$ pse gain & intracellular organelle lumen & \e{1.3}{3}\\
    C $\to$ ana loss & appendage development & \e{4.0}{6}\\
    C $\to$ ana gain & regulation of transcription & \e{7.2}{6}\\
    C $\to$ D loss & gastrulation & \e{9.7}{7}\\
    C $\to$ D gain & mitochondrion & \e{9.5}{5} \\
    D $\to$ amel loss & positive regulation of apoptosis &
    \e{1.1}{2}\\
    D $\to$ amel gain & tissue morphogenesis & \e{7.2}{3} \\
    D $\to$ sim loss & rRNA processing & \e{1.2}{4} \\
    & response to organic substances & \e{3.4}{2}\\
    D $\to$ sim gain & regulation of transcription & \e{4.5}{6}\\
    \hline
  \end{tabular}
  \caption{Functional enrichment for the gene sets gaining or losing
    interactions at each transition state in the phylogenetic tree in
    Figure \ref{fig:phylo}d.}
  \label{tab:trans-enrich}
\end{table}

\begin{table}
  \centering
  \begin{tabular}{|l|l|l|l|}
    \hline
    \textbf{TF} & \textbf{Transition} & \textbf{Functional category} &
    \textbf{\textit{P}-value} \\
    \hline
    BCD & A $\to$ B loss & & \\
    BCD & A $\to$ vir loss & & \\
    BCD, HKB & C $\to$ ana gain & & \\
    BCD, MAD & D $\to$ amel gain & & \\
    DL & B $\to$ per loss & oxidation reduction & \e{4.9}{2}\\
    DL & C $\to$ D gain & mitochondrion & \e{1.2}{8} \\
    MAD & B $\to$ pse gain & & \\
    SLP1 & C $\to$ ana loss & wing disc development & \e{3.5}{5}\\
    & & appendages development & \e{1.8}{5} \\
    & & leg disc pattern formation & \e{2.8}{4} \\
    TWI & B $\to$ pse loss & & \\
    TWI & C $\to$ D loss & gastrulation & \e{5.8}{5} \\
    & & gland development & \e{6.5}{4} \\
    & & tube development & \e{5.2}{3} \\
    \hline
  \end{tabular}
  \caption{Transcription factors significantly enriched ($P<0.05$) for
    targets in gene sets gaining or losing interactions at 
    transition states in the phylogenetic tree in 
    Figure \ref{fig:phylo}d and the functional enrichment of these
    target sets.} 
  \label{tab:trans-tf}
\end{table}

\begin{table}
  \centering
  \begin{tabular}{|l|cccc|}
    \hline
    \textbf{Species} & \textbf{Time points} & \textbf{Series} &
    \textbf{Samples} & \textbf{Completeness} \\
    \hline
    Amel & 	10	&	8	&    56	&	0.7	\\
    Ana	& 	9	&	3	&	27	&  1 \\
    Per	& 	9	&	3	&	27	&  1 \\	 		
    Pse	& 	9	&	3	&	27	&  1\\ 			
    Sim	& 	9	&	3	&	27	&  1	\\ 			
    Vir	& 	13	&	3	&	39	& 0.92 \\
    \hline
  \end{tabular}
  \caption{Expression data summary, listing for each species the
    number of time points, the number of replicate series, the total
    number of samples, and the completeness of the data (number of
    samples divided by number of time points times number of series).}
  \label{tab:data}
\end{table}

\end{document}